 \documentclass[10pt,twocolumn,twoside]{IEEEtran}

\usepackage[pdftex]{graphicx}
\usepackage{cite}
\usepackage[cmex10]{amsmath}
\usepackage{amssymb}  
\usepackage{array}
\usepackage{color}

\newtheorem{theorem}{Theorem}[section]

\newtheorem{definition}{Definition}[section]
\newtheorem{property}{Property}[section]

\newcommand{\comment}[1]{}

\newcommand{\var}[1]{\textbf{var} \left\{ #1 \right\} }
\newcommand{\tr}[1]{\textbf{tr} \left( #1 \right) }

\newcommand{\pgz}{p_g^{(0)}}
\newcommand{\pgo}{p_g^{(1)}}
\newcommand{\pgt}{p_g^{(2)}}
\newcommand{\qgz}{q_g^{(0)}}
\newcommand{\qgo}{q_g^{(1)}}
\newcommand{\qgt}{q_g^{(2)}}
\newcommand{\rgz}{r_g^{(0)}}
\newcommand{\rgo}{r_g^{(1)}}

\newcommand{\bfo}{\textbf{1}}
\newcommand{\Hf}{H_{\text{FO}}}
\newcommand{\Hs}{H_{\text{SO}}}
\newcommand{\vp}{\ensuremath{v}}
\newcommand{\tp}{\ensuremath{m}}
\newcommand{\Pp}{\ensuremath{\overline{P}}}
\newcommand{\Ppg}{\ensuremath{\overline{P}_g}}
\newcommand{\lp}[2]{\ensuremath{\bar{\lambda}_{#1,#2}}}
\newcommand{\lam}[2]{\ensuremath{\lambda_{#1,#2}}}
\newcommand{\pp}[2]{\ensuremath{p_{#1}^{(#2)}}}
\newcommand{\lm}[1]{\ensuremath{\lambda_{#1}}}
\newcommand{\Lg}{\ensuremath{L_{g}}}
\newcommand{\Ng}{\ensuremath{N_{g}}}
\newcommand{\Gg}{\ensuremath{G_{g}}}
\newcommand{\Pg}{\ensuremath{P_g}}
\newcommand{\Ig}{\ensuremath{I_{g}}}
\newcommand{\Qg}{\ensuremath{Q_g}}
\newcommand{\Rg}{\ensuremath{R_g}}
\newcommand{\Sg}{\ensuremath{S_g}}
\newcommand{\Fg}{\ensuremath{F_g}}

\newcommand{\omegnd}[1]{\ensuremath{\Omega_{#1}^{ND}}}
\newcommand{\omegdeg}[1]{\ensuremath{\Omega_{#1}^{D}}}
\newcommand{\gam}[1]{\ensuremath{\Gamma_{#1}^{ND}}}
\newcommand{\gamdeg}[1]{\ensuremath{\Gamma_{#1}^{D}}}
\newcommand{\thet}[1]{\ensuremath{\Theta_{#1}^{ND}}}
\newcommand{\thetdeg}[1]{\ensuremath{\Theta_{#1}^{D}}}

\newcommand{\xti}[1]{\ensuremath{x_{#1}(t)}}

\newcommand{\xot}{\ensuremath{x_1(t)}}
\newcommand{\xtt}{\ensuremath{x_2(t)}}
\newcommand{\xoti}[1]{\ensuremath{x_{1,{#1}}(t)}}
\newcommand{\xtti}[1]{\ensuremath{x_{2,{#1}}(t)}}

\newcommand{\dxot}{\ensuremath{\dot{x}_1(t)}}
\newcommand{\dxtt}{\ensuremath{\dot{x}_2(t)}}

\newcommand{\proj}{\ensuremath{J}}

\begin{document}
%
% paper title
% can use linebreaks \\ within to get better formatting as desired
\title{Consensus and Coherence in Fractal Networks}
%\
%
% author names and IEEE memberships
% note positions of commas and nonbreaking spaces ( ~ ) LaTeX will not break
% a structure at a ~ so this keeps an author's name from being broken across
% two lines.
% use \thanks{} to gain access to the first footnote area
% a separate \thanks must be used for each paragraph as LaTeX2e's \thanks
% was not built to handle multiple paragraphs
%

\author{Stacy~Patterson,~\IEEEmembership{Member,~IEEE} and
        Bassam~Bamieh,~\IEEEmembership{Fellow,~IEEE}% <-this % stops a space
\thanks{S. Patterson is with the Department
of Computer Science, Rensselaer Polytechnic Institute, Troy, NY, 12180, USA, {\tt\small sep@cs.rpi.edu}, 805-455-3457 (contact author)}% <-this % stops a space
\thanks{B. Bamieh is with the Department of Mechanical Engineering, University of California, Santa Barbara,
Santa Barbara, California 93106, USA {\tt\small bamieh@engineering.ucsb.edu}, 805-893-4490}% <-this % stops a space
%\thanks{This works was supported in part by XXXX.}
}

% The paper headers
%\markboth{Journal of \LaTeX\ Class Files,~Vol.~6, No.~1, January~2007}%
%{Shell \MakeLowercase{\textit{et al.}}: Bare Demo of IEEEtran.cls for Journals}
% The only time the second header will appear is for the odd numbered pages
% after the title page when using the twoside option.
% 
% *** Note that you probably will NOT want to include the author's ***
% *** name in the headers of peer review papers.                   ***
% You can use \ifCLASSOPTIONpeerreview for conditional compilation here if
% you desire.

% If you want to put a publisher's ID mark on the page you can do it like
% this:
%\IEEEpubid{0000--0000/00\$00.00~\copyright~2007 IEEE}
% Remember, if you use this you must call \IEEEpubidadjcol in the second
% column for its text to clear the IEEEpubid mark.

% use for special paper notices
%\IEEEspecialpapernotice{(Invited Paper)}

% make the title area
\maketitle

\begin{abstract}

We consider first and second order consensus algorithms in networks with stochastic disturbances. We quantify the deviation from consensus using the notion of network coherence, which can be expressed as an $H_2$ norm of the stochastic system. We use the setting of fractal networks to investigate the question of whether a purely topological measure, such as the fractal dimension, can capture the asymptotics of coherence in the large system size limit. Our analysis for first-order systems is facilitated by connections between first-order stochastic consensus and the global mean first passage time of random walks. We then show how to apply similar techniques to analyze second-order stochastic consensus systems. Our analysis reveals that two networks with the same fractal dimension can exhibit different asymptotic scalings for network coherence. Thus, this topological characterization of the network does not uniquely determine coherence behavior.  The question of whether the performance of stochastic consensus algorithms in large networks can be captured by purely topological measures, such as the spatial dimension, remains open.

\end{abstract}
% IEEEtran.cls defaults to using nonbold math in the Abstract.
% This preserves the distinction between vectors and scalars. However,
% if the journal you are submitting to favors bold math in the abstract,
% then you can use LaTeX's standard command \boldmath at the very start
% of the abstract to achieve this. Many IEEE journals frown on math
% in the abstract anyway.

% Note that keywords are not normally used for peerreview papers.

% For peer review papers, you can put extra information on the cover
% page as needed:
% \ifCLASSOPTIONpeerreview
% \begin{center} \bfseries EDICS Category: 3-BBND \end{center}
% \fi
%
% For peerreview papers, this IEEEtran command inserts a page break and
% creates the second title. It will be ignored for other modes.
\IEEEpeerreviewmaketitle

\section{Introduction}
% The very first letter is a 2 line initial drop letter followed
% by the rest of the first word in caps.
% 
% form to use if the first word consists of a single letter:
% \IEEEPARstart{A}{demo} file is ....
% 
% form to use if you need the single drop letter followed by
% normal text (unknown if ever used by IEEE):
% \IEEEPARstart{A}{}demo file is ....
% 
% Some journals put the first two words in caps:
% \IEEEPARstart{T}{his demo} file is ....
% 
% Here we have the typical use of a "T" for an initial drop letter
% and "HIS" in caps to complete the first word.

\IEEEPARstart{D}{istributed consensus} is a fundamental problem in the
context of multi-agent systems and distributed formation
control \cite{J03,S04}. In these settings, agents must reach
agreement on values like direction, rate of travel, and inter-agent spacing using only local communication,
A critical question is how robust these systems are to external disturbances, and in particular,
how this robustness depends on the network topology. In the presence of stochastic disturbances, a network never reaches consensus, and the best that can be hoped for is that certain measures of deviation from consensus are small. 

In this work, we investigate the performance of systems with first-order and second-order consensus dynamics in the presence of additive stochastic disturbances.  We use an $H_2$ norm as a measure of deviation from consensus, which thus quantifies a notion of \emph{network coherence}. 
For systems with first-order dynamics, it has been shown that this $H_2$ norm can be characterized by the trace of the
pseudo-inverse of the Laplacian matrix~\cite{XBK07,YSL10,BJMP11,lovisari2010resistance}.
This  value has important meaning not just in consensus systems, 
but in electrical networks~\cite{KR93, BBLK94}, random walks~\cite{CRRST97}, and molecular connectivity \cite{GM96}.
For systems with second-order dynamics, this $H_2$ norm is also determined by the spectrum of the Laplacian, though in a slightly 
more complex way~\cite{BJMP11}.  

Several recent works have studied the relationship between network coherence and topology in first-order consensus systems. Young et al.~\cite{YSL10} derived analytical expressions for coherence in rings, path graphs, and star graphs, and Zelazo and Mesbahi~\cite{ZM11} presented an analysis of network coherence in terms of the number of cycles in the graph.  
Our earlier work~\cite{BJMP11} presented an asymptotic analysis of network coherence for both first and second order consensus algorithms in torus 
and lattice networks in terms of the number of nodes and the network dimension.  These results show that there is a marked difference in coherence 
between first-order and second-order systems and also between networks of different spatial dimensions.  For example, in a one-dimensional
ring network with first-order dynamics, the per-node variance of the deviation from consensus grows linearly with the number nodes, 
while in a two-dimensional torus, the per-node variance grows logarithmically with the number of nodes. Even more importantly, 
this work shows that these coherence scalings are the best achievable by any local, linear consensus algorithm.  Thus, in lattice and torus graphs, 
the network dimension imposes a fundamental limitation on the scalability of consensus algorithms.  

A natural question is whether the same dimension-dependent limitations exist in networks with different structures, namely graphs that do not have an integer dimension.
As a first step towards answering this question, we analyze the coherence of first-order and second-order consensus algorithms in
self-similar, tree-like fractal graphs.
For first-order systems, we draw directly from literature on random walks on fractal networks~\cite{LBZ10,ZWZZGW10} to
show that, in a network with $N$ nodes, the network coherence scales as $N^{1/d_f}$ where $d_f$ is the \emph{fractal dimension} (also the Hausdorff dimension, in our case) of
the network. 
We then show how the techniques used for the analysis of random walks  can be extended to analyze the coherence of second-order consensus systems, and we present asymptotic results for the
 per-node variance in terms of the network size and fractal dimension.  An interesting, and perhaps unexpected result of our analysis is that the fractal dimension does not uniquely determine the asymptotic behavior of network coherence. We show that two self-similar graphs with the same fractal dimension exhibit different coherence scalings.   We note that a preliminary version of this work, without mathematical derivations, appeared in~\cite{PB11}.

The remainder of this paper is organized as follows.   In Section~\ref{coherence.sec}, we
present the models for first-order and second-order noisy consensus systems and give a formal
definition of network coherence for each setting.  We also present several properties from other 
domains that are mathematically similar to network coherence.  In Section~\ref{graphs.sec},
we describe the fractal graph models, and in Section~\ref{fractals.sec} we present 
analytical results on the  coherence scalings for these fractal graphs.  
Section~\ref{dimension.sec} presents a discussion of the relationship between graph dimension and coherence, followed 
by our conclusion in Section~\ref{conclusion.sec}.

\section{NETWORK COHERENCE}
\label{coherence.sec}
We consider local, linear first-order and second-order consensus algorithms over an undirected, connected
network modeled by an undirected graph $G$ with $N$ nodes and $M$ edges. 
We denote the adjacency matrix of $G$ by $A$, and $D$ is a diagonal matrix where the diagonal entry $D_{i,i}$ is
the degree of node $i$.  The Laplacian matrix of the graph $G$ is denoted by $L$ and is defined as $L: = D - A$.

Our objective is to study the robustness of consensus algorithms when the nodes are subject to external perturbations and to analytically quantify the relationship between the system robustness and the graph topology.
We capture this robustness using a quantity that we call \emph{network coherence}.
We now formally define the system dynamics and the notion of network coherence (Sections~\ref{firstdef.sec} and \ref{seconddef.sec}).
We then present some mathematically related properties that we can leverage in our robustness analysis (Section~\ref{relprop.sec}). 

\subsection{Coherence in Networks with First-Order Dynamics} \label{firstdef.sec}

In the first-order consensus problem, each node $j$ has a single state $\xti{j}$.  The state of the entire system at time $t$ is given by the vector $x(t) \in \mathbb{R}^N$. 
Each node state is subject to stochastic disturbances, and the objective is for the nodes to maintain consensus at the average of their current states.

The dynamics of this system are given by,
\begin{equation}
\dot{x}(t) = -\beta L x(t) + w(t),
\label{fo.eq}
\end{equation}
where $\beta$ is the gain on the communication links, and $w(t)$ is a size $N$ disturbance vector with zero-mean, unit variance, and uncorrelated second-order processes. 

In the absence of the disturbance processes, the system converges asymptotically to consensus at the
the average of the initial states~\cite{T84}.
With the additive noise term, the nodes do not converge to consensus, but instead, node values fluctuate
around the average of the \emph{current} node states.   

The concept of network coherence captures the variance of these fluctuations in the first-order consensus system.
\begin{definition}The \emph{first-order network coherence} is defined  as the mean (over all nodes), steady-state variance of
the deviation from the average of the current node states,
\[
\Hf :=  \lim_{t \rightarrow \infty}  \frac{1}{N}  \sum_{j=1}^{N}\var{\xti{j} - \frac{1}{N}\sum_{k=1}^N \xti{k}}.
\]
\end{definition}

We define the output of the system (\ref{fo.eq}) to be
\begin{equation}
y(t) = \proj x(t),
\label{output.eq}
\end{equation}
where $\proj$ is the projection operator $\proj := I - \frac{1}{N} \bfo \bfo^*$, with $\bfo$ the $N$-vector of all ones.
It is well known that $\Hf$ is given by the $H_2$ norm of the system defined in  (\ref{fo.eq}) and (\ref{output.eq}),
\[
\Hf = \frac{1}{N} \tr{\int_0^{\infty} e^{-\beta L^*t}\proj e^{-\beta Lt}dt}.
\]
It has been shown that $\Hf$ is s completely determined by the spectrum of $L$~\cite{XBK07,BJMP11,YSL10}.
Let the eigenvalues of $L$ be denoted $0 = \lm{1} < \lm{2} \leq \ldots \leq \lm{N}$.
The first-order network coherence is then equal to,
\begin{equation}
\Hf = \frac{1}{2\beta N} \sum_{i=2}^N \frac{1}{\lm{i}} .
\label{focoh.eq}
\end{equation}

\subsection{Coherence in Networks with Second-Order Dynamics} \label{seconddef.sec}
In the second-order consensus problem, each node $j$ has two state variables $\xoti{j}$ and $\xtti{j}$.
The state of the entire system is thus captured in two $N$-vectors, $\xot$ and $\xtt$.
Nodes update their states based on local feedback, i.e., the states of their neighbors in the graph, and they 
are also subject to random external disturbances that enter through the $\xtt$ terms.
The dynamics of the system are,
\begin{equation}
\left[\begin{array}{c}
\dxot \\
\dxtt
\end{array} \right] = \left[\begin{array}{rr}
0 & I \\
-\beta L & -\beta L 
\end{array} \right] \left[\begin{array}{c}
\xot \\
\xtt
\end{array} \right] +\left[\begin{array}{c}
0 \\
I
\end{array} \right] w(t),
\label{do.eq}
\end{equation}
where $w(t)$ is a $2N$ disturbance vector with zero-mean, unit variance, and uncorrelated second-order processes.

These system dynamics arise in the problem of autonomous vehicle formation control (e.g., see~\cite{BJMP11}).  
Here, $\xot$ is an $N$-vector containing the vehicles' positions and $\xtt$ is an $N$-vector containing the vehicles' velocities. 
The vehicles attempt to maintain a specified formation traveling at a fixed velocity while subject to stochastic external perturbations.  
The non-zero entries of $L$ specify the communication links of the formation, i.e., if $L_{ij} = -1$,
then node $i$ can observe the position and velocity of node $j$, and vice versa.   
It has also been shown that similar system dynamics also arise in problems in phase synchronization in power networks~\cite{BP13}.

The network coherence of the second-order system (\ref{do.eq}) is defined in terms $\xot$ only, and as with first-order coherence,
it captures the deviation from the average of $\xot$.
\begin{definition}
The \emph{second-order network coherence} is the mean (over all nodes), steady-state variance of the
deviation from the average of $\xot$,
\[
\Hs := \frac{1}{N} \sum_{j=1}^{N} \lim_{t \rightarrow \infty} \var{\xoti{j} - \frac{1}{N} \sum_{k=1}^N \xoti{k}}.
\] 
\end{definition}
In the vehicle formation problem, this quantity captures the deviation of the vehicle positions from a rigid formation traveling at the average position.

We define the output for the system (\ref{do.eq}) as
\begin{equation}
y(t) =  \left[\begin{array}{cc} \proj & 0\end{array}\right]  \left[\begin{array}{c}
\xot \\
\xtt
\end{array} \right],
\label{dooutput.eq}
\end{equation}
where $\proj $ is again the projection operator, $\proj := I - \frac{1}{N} \bfo \bfo^*$.
The second-order network coherence is given by the 
$H_2$ norm of the system defined by (\ref{do.eq}) and (\ref{dooutput.eq}).  This value is
also completely determined by the eigenvalues of the Laplacian matrix~\cite{BJMP11}, specifically, 
\begin{equation}
\Hs = \frac{1}{2\beta^2 N} \sum_{i=2}^N \frac{1}{\left(\lm{i}\right)^2}.
\label{socohdef.eq}
\end{equation}
\subsection{Related Concepts} \label{relprop.sec}
\label{concepts.sec}
The eigenvalues of the Laplacian are linked to the topology of the network, and therefore, it is not surprising that
these eigenvalues play a role in many graph problems.  In fact, the sum,
\begin{equation} \label{S.eq}
S := \sum_{i=2}^{N} \frac{1}{\lm{i}}
\end{equation}
that appears in the expression for first-order coherence in (\ref{focoh.eq}) is an important quantity, not just in the study
of consensus algorithms, but in several other fields. We can leverage work in these fields to develop analytic expressions for
 network coherence for different graph topologies. We briefly review these related properties below.

\subsubsection{Effective resistance in an electrical network}
Let the graph represent an electrical network where each edge is a unit resistor.
The \emph{resistance distance} $r_{ij}$ between two nodes $i$ and $j$ is the potential distance between
them when a one ampere current source is connected from node $j$ to node $i$.
The \emph{total effective resistance} of the network, also called the \emph{Kirchoff index} \cite{KR93, BBLK94}, is 
the sum of  the resistance distances over all pairs of nodes in the graph.
It has been shown~\cite{AF02} that the total effective resistance depends on the spectrum of the Laplacian matrix as,
\[
R = 2 N S = 2N \sum_{i=2}^{N} \frac{1}{\lm{i}}.
\]

\subsubsection{Global mean first passage Time of a random walk}
In a simple random walk  on a undirected graph, the probability of moving from a node $i$ to a neighboring node $j$ is $\frac{1}{d_i}$ where $d_i$ 
is the out-degree of node $i$.  The \emph{first passage time} $f_{ij}$ is the average number of steps it takes
for a random walk starting at node $i$ to reach node $j$ for the first time.  The \emph{global mean first passage time} is the average first passage
time over all pairs of nodes.
It has been shown that, for a connected graph, the mean first passage time between nodes $i$ and $j$ and the resistance distance are related as
$f_{ij} + f_{ji} = 2M r_{ij}$, where $M$ is the number of edges in the graph~\cite{CRRST97}.  The global mean first passage time is therefore related to the effective resistance as 
\begin{equation} \label{gmfp.eq}
F = \frac{M}{N(N-1)}R =  \frac{2M}{N-1} S. 
\end{equation}
If the graph is a tree, then $M=N-1$, and the global mean first passage time is simply $F = 2 S$.

\subsubsection{Quasi-Wiener index}
The Wiener index is a measure of molecular connectivity~\cite{W47}. Here, the graph represents a molecule where nodes are atoms and edges are chemical bonds, and the distance between two atoms is the length (number of edges) of the shortest path between them.
The Wiener index is the sum of the distances between
all pairs of non-hydrogen atoms.  If the molecular graph is acyclic, this value is exactly $S$ in (\ref{S.eq})~\cite{GM96}.  If the graph contains cycles, the Wiener index is no longer equal to the sum of lengths of the shortest paths.  However, this quantity is still
utilized in mathematical chemistry and is called the quasi-Wiener index~\cite{MGB95}.

\section{SELF-SIMILAR GRAPHS}
\label{graphs.sec}
Ideally, one would like to find an analytical expression for network coherence that depends on the graph topology.
While it is a difficult problem to characterize the spectrum of the Laplacian matrix for a general graph,
for graphs with special structure, it is sometimes possible to find a closed form for the either the eigenvalues themselves or for
the sum of their inverses.  For example, for $d$-dimensional torus and lattice networks, the Laplacian is a circulant operator.
Its eigenvalues can be determined analytically using a Discrete Fourier Transform, and analytical expressions that relate coherence to the network size and dimension have been derived~\cite{M69,BH08,BJMP11}.

%We note that results similar to those for $\Hf$ has also appeared in the contexts of global mean first passage time \cite{M69} and effective resistance \cite{BH08} in lattice and torus networks.

To extend this type of analysis to other graph topologies, we require that such graphs have both a dimension and a prescribed method of increasing the graph size that preserves this dimension.
A class of graphs that exhibits these properties is the class of self-similar graphs.
Informally, a self-similar graph is one which exhibits the same structure at every scale.  For a more formal definition, we refer the reader to~\cite{K04}.
One notion of dimension of a self-similar graph is the \emph{fractal dimension}, which is defined as follows~\cite{T90}.
\begin{definition}
Let $G = (V,E)$  be an infinite, connected, undirected graph where each vertex has finite degree. 
Let $v \in V$ be an arbitrary vertex, and define $B(r)$ to be the of radius $r$, centered at $v$, i.e.,
\[
B(r) = {u \in V: d(u,v) \leq r},
\]
where $d(u,v)$ denotes the length (number of edges) in the shortest path between $u$ and $v$ in $G$.
The \emph{fractal dimension} of $G$ is
\[
d_f := - \limsup_{r \rightarrow \infty} \frac{\ln(B(r))}{\ln(r)}.
\]
\end{definition}

Other notions of dimensions include the Hausdorff dimension and the box counting dimension.
For self-similar graphs of the type we study in this work, these values of these three dimensions are equivalent~\cite{K04}.

Torus and lattice graphs are both self-similar graphs, and their fractal dimensions are equivalent 
to the natural dimension definition, e.g., a 2-dimensional torus has $d_f = 2$.
In this work, we give coherence analysis for two classes of self-similar graphs that have fractional dimensions,  tree-like fractals and Viscek fractals.  
We now describe the construction of these graphs.

\subsubsection*{Tree-Like Fractals}
%
%\begin{figure}
%    \centering
%    \begin{minipage}[b]{0.45\linewidth}
%	\centering
%        \includegraphics[width=.85\textwidth]{basin2}
%        \caption{First four generations of the tree-like fractal for $\tp=2$.}
%     \label{basin.fig}
%   \end{minipage}
%   ~~~
%\begin{minipage}[b]{0.45\linewidth}
%\centering
%        \includegraphics[width=.85\textwidth]{vicsek2}
%        \caption{First three generations of the Vicsek fractal for $\vp=4$ \note{change $g=0$ to $g=1$}.}
%        \label{vicsek.fig}
%  \end{minipage}
%  \vspace{-.4cm}
%\end{figure}

\begin{figure}
\centering
\includegraphics[scale=.45]{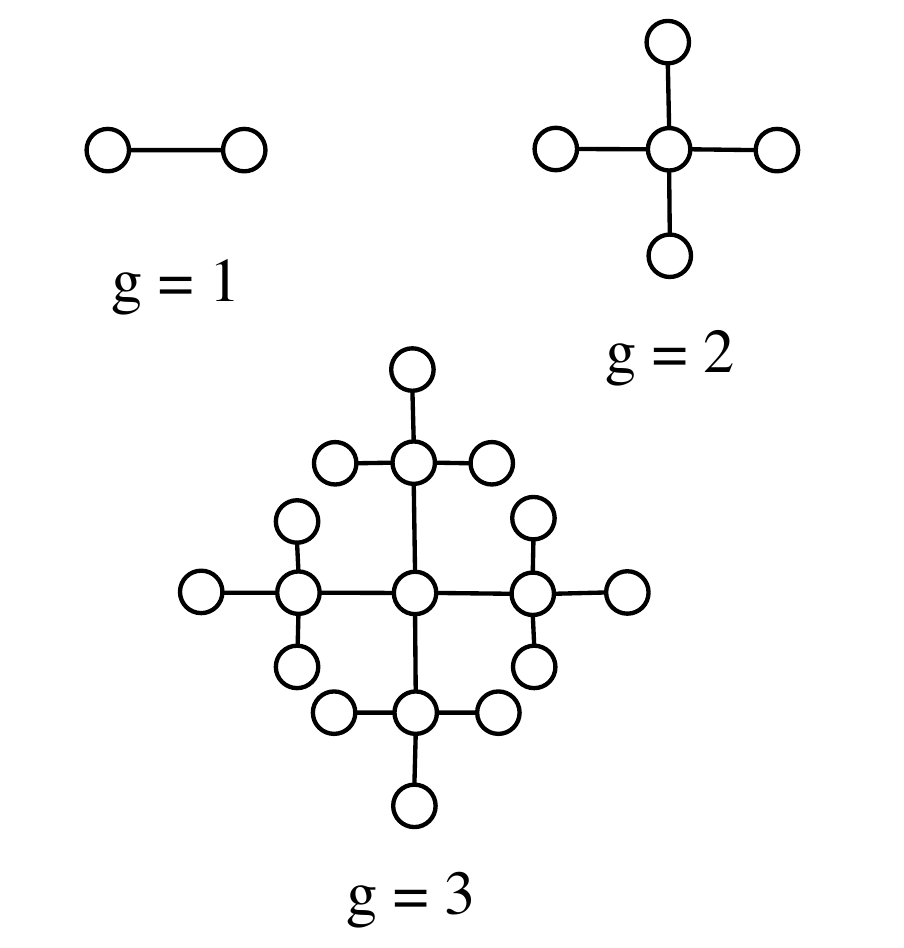}
\caption{First three generations of the tree-like fractal for $\tp=2$.}
\label{basin.fig}
\end{figure}

\begin{figure}
\centering
\includegraphics[scale=.37]{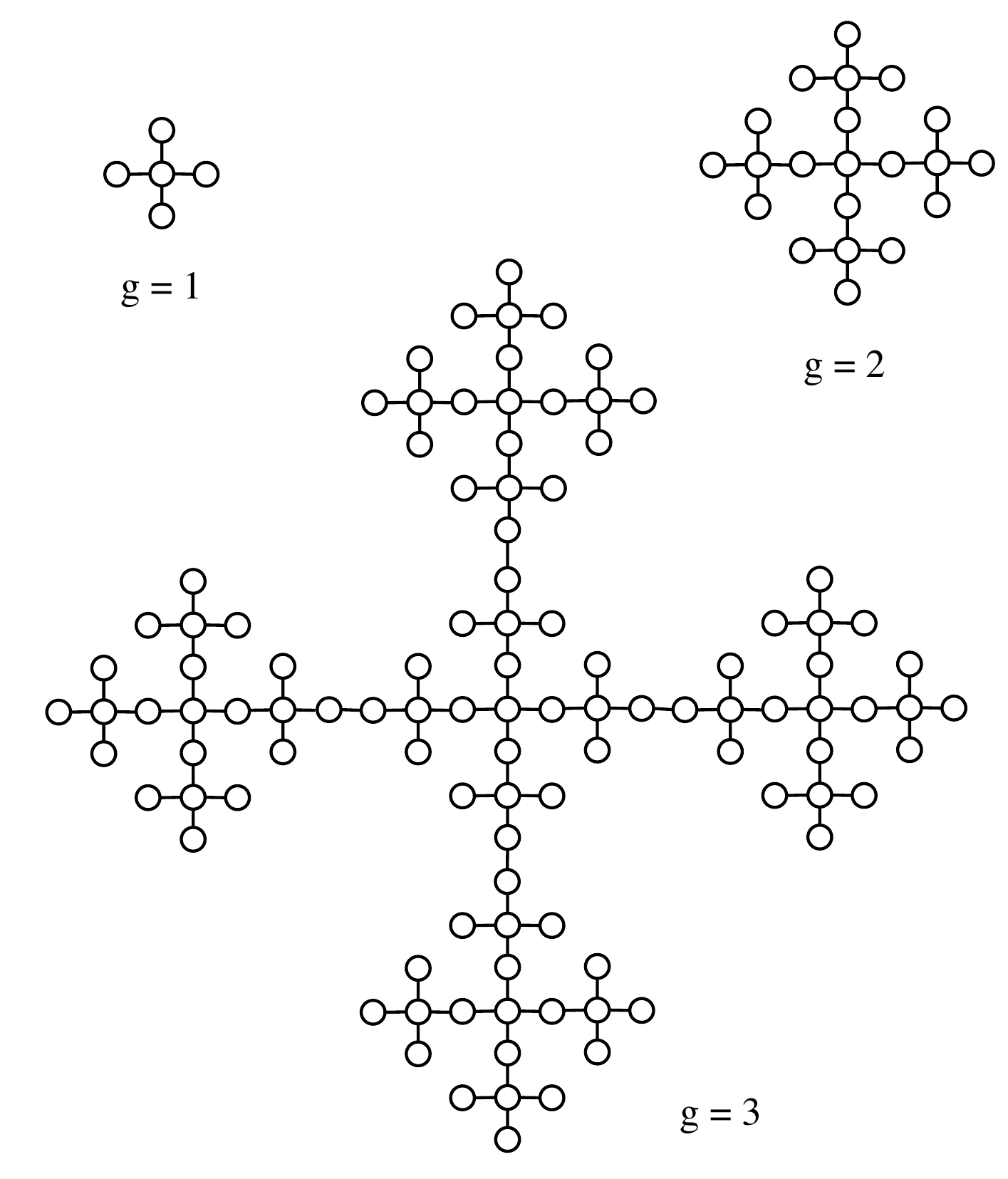}
\caption{First three generations of the Vicsek fractal for $\vp=4$.}
\label{vicsek.fig}
\end{figure}

Each family of tree-like fractal graphs is parameterized by a positive integer $\tp$.  
The graphs are constructed in an iterative manner, and each iteration yields a new graph generation.   
This generation $1$ graph consists of two vertices, or nodes, connected by a single edge.
  Given a graph of generation $g$, denoted $\Gg$, the graph of generation $g+1$ is formed by 
replacing each edge with a path of length 2.  In other words, each edge $(i,j)$ in $\Gg$,
is replaced by two edges $(i,k)$ and $(k,j)$ where $k$ is a new node (not existing in graph $\Gg$).  
Then, $\tp$ additional new nodes are added to every new node $k$.  The generation $g$ graph thus has $(\tp+2)^g + 1$ nodes.
The process is illustrated in Fig. \ref{basin.fig} for $\tp = 2$.  
This tree-like fractal model encompasses several
well-known fractal graphs, including the T-graph  ($\tp = 1$)~\cite{A08} and the Peano basin fractal ($\tp = 2$)~\cite{DDV09}.
The tree-like fractal graph has a fractal dimension of $d_f = \log(\tp+2) / \log(2)$~\cite{LBZ10}.

% and spectral dimension $d_s = 2 \log(k+2) / \log(2(k+2))$.

\subsubsection*{Vicsek Fractals}
The family of Vicsek fractal graphs is also parameterized by a positive integer $\vp$, and each family is constructed in an iterative manner~\cite{V83,BFK04}.
The generation 1 graph is a star graph with $\vp + 1$ nodes. The graph for generation $g+1$ is generated from the generation $g$
graph by making $\vp$ copies of $\Gg$ and arranging them in a star around $\Gg$.
These copies are connected to $\Gg$ by adding edges from the $\vp$ corners of the $\Gg$, each one linking to a corner of 
a copy. Thus, the generation $g$ graph has $\Ng = (\vp+1)^g$ nodes.  We illustrate the first three generations of the Vicsek fractal for $\vp=4$ in Fig. \ref{vicsek.fig}.   
The Vicsek fractal has a fractal dimension of
$d_f = \log(\vp+1) / \log(3)$. 
% and spectral dimension $d_s = 2 \log(f+1) / \log(3f + 3)$.

\section{COHERENCE ANALYSIS}
\label{fractals.sec}

In this section, we present our analysis of network coherence for the families of fractal graphs described in the previous section.  
Several recent works have analyzed the global mean first passage time of a random walk for these graph families~\cite{BFK04,ZWZZGW10,LBZ10}.
Using the relationship described in Section \ref{concepts.sec}, we can extend this analysis to derive asymptotic scalings for network coherence in systems with first-order noisy consensus dynamics.  
Below, we briefly summarize the analytical techniques used in previous works and formally state the asymptotic scalings for first-order network coherence in fractal networks.
We then derive expressions for network coherence in fractal networks with second-order dynamics.

\subsection{Coherence in Tree-Like Fractals} \label{treecoherence.sec}
In their recent work, Lin~et~al. analyze the global mean first passage time in tree-like fractals~\cite{LBZ10}.
Their approach is based on first deriving a recursion for the characteristic polynomial of the Laplacian matrix.
They then show that the sum $S$ in (\ref{S.eq}) can be obtained by solving for several 
coefficients of this characteristic polynomial.  
We first review the construction of this polynomial and then show how it can be used to derive expressions for 
network coherence.

Let $\Gg$ be the graph of generation $g$ with $\Ng$ nodes, and let $\Lg$ denote its Laplacian. 
The characteristic polynomial for $\Lg$ is
\begin{equation}
\Pg(x) = \det(\Lg - x \Ig).
\label{char.eq}
\end{equation}
Here $\Ig$ is the $\Ng \times \Ng$ identify matrix. As we are only interested in the non-zero eigenvalues of $\Lg$, we instead consider a modified
version of the characteristic polynomial,
\begin{equation}
\Ppg(x) = \frac{1}{x} \Pg(x).
\label{charmod.eq}
\end{equation}
The modified characteristic polynomial $\Ppg$ can be written as,
\begin{equation}
\Ppg(x) = \sum_{i=0}^{\Ng-1} \pp{g}{i} x^i =  \pp{g}{\Ng-1} \prod_{i=1}^{\Ng-1} (x- \lp{g}{i}),
\label{viete.eq}
\end{equation}
where $\pp{g}{i}$ denotes the coefficient of the term $x^i$ and $\lp{g}{i}$, $i=1 \ldots (\Ng-1)$, are the roots of $\Ppg(x)$.

In the remainder of Section~\ref{treecoherence.sec}, we show how to determine network coherence from coefficients of (\ref{viete.eq}).

\subsubsection{First-order network coherence}\label{treefirst.sec}

For systems first-order dynamics, network coherence depends on the sum of the roots of (\ref{charmod.eq}),
\[
\Sg = \sum_{i=2}^{\Ng} \frac{1}{\lam{g}{i}} = \sum_{i=1}^{\Ng-1} \frac{1}{\lp{g}{i}},
\]
where $\lam{g}{i}$, $i =1 \ldots \Ng$ are the roots of $\Pg(x)$.
Lin~et~al. show that the sum $\Sg$ can be expressed in terms of the zeroth order and first order coefficients of (\ref{viete.eq})~\cite{LBZ10},
\begin{equation}
\label{foco.eq}
\Sg = - \frac{\pgo}{\pgz}.
\end{equation}
Therefore, to find this sum, one only needs to determine these two coefficients.

The coefficients can be found by first deriving a recursion for $\Ppg(x)$, as follows.  Let 
$\Qg$ be the characteristic polynomial of the $(\Ng-1) \times (\Ng-1)$ submatrix of $\Lg$ formed by removing a single column and row
corresponding to an outermost node.  Let $\Rg$  be the characteristic polynomial of the $(\Ng-2) \times (\Ng-2)$ submatrix of $\Lg$ formed by removing columns and rows 
corresponding to two outermost nodes.
The following equations capture the relationship between the modified  characteristic polynomials for $\Lg$ and $L_{g+1}$,
\begin{align}
\Pp_{g+1}(x)=&(m+2)[ \Qg(x)]^{m+1}\Ppg(x) + (m+1) [ \Qg(x)]^{m+2} \label{P.eq} \\
Q_{g+1}(x)=&[ \Qg(x)]^{m+2} + (m +1) x \Rg(x) [ \Qg(x)]^{m +1}  \nonumber \\
& + (m +1) x \Rg(x)[ \Qg(x)]^m \Pp_g(x) \label{Q.eq}) \\
R_{g+1}(x)=& 2\Rg(x)[\Qg(x)]^{m+1} + (m+1)x[ \Rg(x)]^2 [\Qg(x)]^m \nonumber \\
&+ m x [\Rg(x)]^2 [\Qg(x)]^{m-1} \Ppg(x). \label{R.eq}
\end{align}

To find $\pgz$, first let $\qgz$ and $\rgz$ be the constant terms for $\Qg(x)$ and $\Rg(x)$ respectively (when written in the form like (\ref{viete.eq})).
One can then write recursions on these constant terms as,
\begin{eqnarray*}
p_{g+1}^{(0)} &=& (m+2)[\qgz]^{m+1}\pgo + (m+1) [\qgz]^{m+2} \\
q_{g+1}^{(0)} &=& [\qgz]^{m+2} \\
r_{g+1}^{(0)} &=& 2 \rgz [\qgz ]^{m+1}.
\end{eqnarray*}
Given the initial values for the generation 1 graph, $p_{1}^{(0)}=-m-3$,  $q_1^{(0)} = 1$, and $r_1^{(0)} = 2$, one can solve for $\pgz$, $\qgz$, and $\rgz$.
Using a similar approach, one can also solve for $\pgo$ and, therefore, for the entire expression $\Sg$ (see ~\cite{LBZ10} for details).

With these results, Lin~et~al. arrive at the following expression for the asymptotic order of
the  global mean first passage time in a tree-like fractal of generation $g$ with $\Ng$ nodes,
\[
\Fg  = 2 \sum_{i=2}^{\Ng}  \frac{1}{\lam{g}{i}}  \sim \Ng^{1 + \log(2)/\log(m+2)}.
\]

By using the relationship between $\Fg$ and $\Sg$ defined in Section \ref{concepts.sec} ,
we can easily obtain an analytical expression for the coherence of first-order consensus algorithms in tree-like fractals.
\begin{theorem}
For a tree-like fractal with $N$ nodes, parameterized by the integer $m$, the first-order network coherence of the system with dynamics defined in (\ref{fo.eq}) is given by
\[
\Hf \sim  \frac{1}{\beta} N^{\log(2) / \log(m+2)} = \frac{1}{\beta} N^{1/d_f},
\]
where $d_f$ is the fractal dimension of the graph.
\end{theorem}

\subsubsection{Second-order network coherence}\label{treesecond.sec}

We now show how we extend the work above to derive an expression for second-order network coherence in tree-like fractals.
In this case, for the generation $g$ graph, we are interested in a sum of the form,
\begin{equation} \label{sumsq.eq}
\sum_{i=1}^{\Ng-1} \frac{1}{\left(\lp{g}{i}\right)^2}.
\end{equation}
Using equation (\ref{viete.eq}) and Vieta's formulae, we can express this sum in terms of coefficients of $\Pp(x)$.
For a tree-like fractal of generation $g$ with $\Ng$ nodes, the sum is,
\begin{equation}
\label{doco.eq}
\sum_{i=1}^{\Ng-1} \frac{1}{\left(\lp{g}{i}\right)^2} = \left( \frac{\pgo}{\pgz}\right)^2 - 2 \frac{\pgt}{\pgz},
\end{equation}

The values for $\pgz$ and $\pgo$ were derived by Lin~et~al. in their analysis of the global mean first passage time.
What remains is to solve for $\pgt$.
To do this, we first find the recursion equations for the coefficients corresponding to the first-order term in $\Rg$ in (\ref{R.eq}) and the 
second-order terms in $\Ppg$ and $\Qg$ in (\ref{P.eq}) and (\ref{Q.eq}), respectively. 
\begin{align}
&r_{g+1}^{(1)}=2 [\qgz]^{\tp+1} \rgo + 2(\tp+1) \rgz [\qgz]^\tp \qgo \nonumber \\
&~~+ (\tp+1) [\rgz]^2[\qgz]^\tp + \tp [\rgz]^2 [\qgz]^{\tp-1} \pgz \label{rectree1r.eq}\\
&p_{g+1}^{(2)}=(m+2)[\qgz]^{m+1}\pgt \nonumber\\
&~~+ (m+1)(m+2)[\qgz]^m\qgt \pgz \nonumber\\
&~~+\textstyle \frac{m(m+1)(m+2)}{2}[\qgz]^{m-1}[\qgo]^2 \pgz\nonumber \\
&~~+ (m+1)(m+2)[\qgz]^m\qgo\pgo \nonumber \\
&~~+(m+1)(m+2)[\qgz]^{m+1} \qgt \nonumber\\
&~~+\textstyle\frac{(m+1)^2(m+2)}{2} [\qgz]^m [\qgo]^2. \label{rectree2p.eq} \\
&q_{g+1}^{(2)}= (m+2) [\qgz]^{m+1}\qgt+\textstyle \frac{(m+1)(m+2)}{2}[\qgz]^m[\qgo]^2 \nonumber \\
&~~+(m+1)\rgo [\qgz]^{m+1} + (m+1)^2 \rgz [\qgz]^m \qgo \nonumber \\
&~~+ (m+1)[\qgz]^m \pgz \rgo+m(m+1) \rgz \pgz [\qgz]^{m-1}\qgo \nonumber \\
&~~+ (m+1) \rgz [\qgz]^m \pgo  \label{rectree2q.eq}
\end{align}
We then solve for these coefficients.  The full derivation is given in Appendix~\ref{tree.app}.

As we are interested in the asymptotic behavior of $\Hs$, we consider only the highest order terms of the coefficients in (\ref{doco.eq}),
which are given by,
\begin{align*}
\pgz &\sim -(\tp+2)^g\\
\pgo &\sim 2^g(\tp+2)^{2g} \\
\pgt &\sim - 2^{2g}(\tp+2)^{3g}.
\end{align*}
Thus,  the order of the sum in (\ref{sumsq.eq}) is,
\[
\sum_{i=1}^{\Ng-1} \frac{1}{\left(\lp{g}{i}\right)^2} \sim 2^{2g}(\tp +2)^{2g} \sim \Ng^{2 + 2\log(2)/\log(\tp+2)}.
\]
Here, the last expression follows from the fact that the generation $g$ tree-like fractal graph has $\Ng = (m+2)^g + 1$ nodes.

We then substitute the expression for this sum into $\Hs$ in (\ref{socohdef.eq}) to arrive at the following theorem.
\begin{theorem}
For a tree-like fractal with $N$ nodes, parameterized by the integer $m$, the second-order coherence of the system with dynamics as defined in (\ref{do.eq}) is given by
\[
\Hs \sim \frac{1}{\beta^2}N^{1 + 2 \log(2) / \log(m+2)} = \frac{1}{\beta^2}N^{1 + (2/d_f)},
\]
where $d_f$ is the fractal dimension of the graph.
\end{theorem}

\subsection{Generalized Vicsek Fractals} 
To analyze the coherence of consensus algorithms in Vicsek fractals, we exploit a different technique that was used in the analysis of the global mean first passage time.
As with tree-like fractals, it is not straightforward to find a closed form for the individual eigenvalues of the Laplacian matrix of a Vicsek fractal.  However, 
in their recent work~\cite{ZWZZGW10}, Zhang et al. determined a closed-form expression for the sum
$S$ in (\ref{S.eq}).  With this sum, we can easily obtain the first-order network coherence.  
We first state the results for first-order coherence and then show how we can use a similar approach to obtain a closed-form expression for second-order coherence.  

This analysis makes use of several previously derived properties relating to the eigenvalues of $\Lg$ for Vicsek fractals~\cite{JWC92,JW94,BFK04}, which we state here convenience.
\begin{property} \label{vicsek.prop}
Let $\Lg$ be the Laplacian matrix for a generation $g$ Vicsek fractal.  The eigenvalues of $\Lg$ satisfy the following: 
\begin{enumerate}
\item The non-degenerate eigenvalues of $L_1$ are 0 and $\vp+1$.  $L_1$ has one degenerate eigenvalue with value one and multiplicity $\vp-1$.
\item Every eigenvalue of $\Lg$ is also an eigenvalue of $L_{g+1}$.  As a result, eigenvalues preserve their degeneracy in subsequent generations.
\item For $\Lg$, the multiplicity of the one eigenvalue is,
 \begin{equation}
 \Delta_g := (\vp-2)(\vp+1)^{g-1} + 1.  \label{delta.eq}
\end{equation} 
 A degenerate eigenvalue that appears for the first time at generation $j$ has multiplicity $\Delta_{g-j}$. 
\item Each non-zero eigenvalue $\lam{g}{i}$ in $\Lg$ produces three new eigenvalues in $L_{g+1}$ according the relation,
\begin{equation}
 \lam{g+1}{i}(\lam{g+1}{i}- 3) (\lam{g+1}{i} - \vp - 1) = \lam{g}{i}.
\label{vicrec.eq}
\end{equation}
\end{enumerate}
\end{property}

\subsubsection{First-order network coherence}
\label{vicsekfirst.sec}
To find the sum $\Sg$, Zhang~et~al.~\cite{ZWZZGW10} consider the sums for degenerate and non-degenerate eigenvalues separately.
 Let $\omegnd{g}$ be the set of non-degenerate eigenvalues of $\Lg$, excluding 0, and let $\omegdeg{g}$ be the set of degenerate eigenvalues.
Thus $\Sg$ is equivalent to,
\[
\Sg = \sum_{\lambda \in \omegnd{g}} \frac{1}{\lambda} + \sum_{\lambda \in \omegdeg{g} } \frac{1}{\lambda}.
\]

In the generation $1$ graph, there is a single  non-degenerate, non-zero eigenvalue, $\vp+1$.
This eigenavalue produces three non-degenerate eigenvalues in generation 2, according to (\ref{vicrec.eq}).   
These eigenvalues are the first-generation \emph{descendants} of $\vp+1$. 
The first generation descendants yield $3^2$ second-generation descendants (also non-degenerate), and so on.  
Let $\gam{i}$ be the sum of the reciprocals of the $i^{th}$ generation descendants of $\vp+1$. 
By employing the recursion on the eigenvalues defined by (\ref{vicrec.eq}) in Property \ref{vicsek.prop}, 
 Zhang et al. obtain the following closed-form expression for $\gam{i}$,
\begin{equation}
\gam{i} =  \sum_{\lambda \in (\omegnd{i} - \omegnd{i-1})} \frac{1}{\lambda} = 3^i(\vp+1)^{i-1}. \label{gamsum1.eq}
\end{equation}
They then sum over the generations of descendants, $i=0 \ldots g-1$,  to find the sum of
the inverses of the non-degenerate eigenvalues of $\Lg$,
\begin{equation}
 \sum_{\lambda \in \omegnd{g}} \frac{1}{\lambda} = \sum_{i=0}^{g-1} \gam{i} = \frac{1}{\vp+1} \frac{3^g(\vp+1) ^g - 1}{3 \vp + 2}. \label{ndsum.eq}
 \end{equation}

A similar approach can be used to find a closed-form expression for the sum of reciprocals of  the $i^{th}$ generation descendants of a single degenerate eigenvalue $1$,
\begin{equation} \label{degeclosed.eq}
\gamdeg{i} = 3^i (v+1)^i.
\end{equation}
This expression can then be combined with the multiplicity of the degenerate eigenvalues defined in Property \ref{vicsek.prop} to find the sum of the inverses of the degenerate eigenvalues of $\Lg$,
\begin{align}
&\sum_{\lam{g}{i} \in \omegdeg{g} } \frac{1}{\lam{g}{i}} = \sum_{i=0}^{g-1} \Delta_{g-i} \gamdeg{i} \nonumber \\
&= \frac{1}{2}(v-2)(v+1)^{g+1}(3^g - 1)+ \frac{2(v+2)}{v+1} \frac{3^g(v+1)^g-1}{3v + 2}. \label{gamsum2.eq}
\end{align}

It is then straightforward to add  (\ref{gamsum1.eq}) and (\ref{gamsum2.eq}) to obtain $\Sg$,
\[
\Sg = \frac{(v-2)(v+1)^{g-1}(3^g-1)}{2} + \frac{v+2}{v+1} \frac{3^g(v+1)^g - 1}{3v+2}.
\]
Using the facts that the number of nodes in generation $g$ graph is $\Ng = (\vp+1)^g$ and that $3^g = \Ng^{\log(3)/\log(\vp+1)}$,
the sum $\Sg$ can be shown to be of the following order,
\[
\Sg \sim \Ng^{1 + \log(3) / \log(\vp+1)} = \Ng^{1 + 1 / d_f}.
\]

The expression for network coherence in Vicsek fractals with first-order consensus dynamics immediately follows from $\Sg$ and is formally stated in the following theorem. 
\begin{theorem}
For a generalized Vicsek fractal graph with $N$ nodes, parameterized by the positive integer $\vp$, the first-order coherence of the system with the dynamics defined in (\ref{fo.eq}) is given by
\[
\Hf \sim  \frac{1}{\beta} N^{\log(3)/\log(\vp+1)} =  \frac{1}{\beta} N^{1 / d_f},
\]
where $d_f$ is the fractal dimension of the graph.
\end{theorem}

\subsubsection{Second-order network coherence}
\label{vicseksecond.sec}
To find the second-order network coherence, we also determine a recursion over the eigenvalues of $\Lg$, in this case, a recursion over the sum of the squares of the
inverses of the eigenvalues.  

As the first step in this analysis, we consider the relationship between the eigenvalues of $\Lg$ and $L_{g+1}$.
Let $\lam{g}{i}$ be a non-zero eigenvalue of $\Lg$.   This eigenvalue produces three new eigenvalues in $L_{g+1}$, according to (\ref{vicrec.eq}),
which we denote by  $\lam{g+1}{i_1}$, $\lam{g+1}{i_2}$, and $\lam{g+1}{i_3}$. 
We want to define the sum of the squared inverses of these three eigenvalues in terms of the parent eigenvalue.

We first expand the expression for the sum of squared inverses, as follows, 
\begin{align}
&\textstyle\frac{1}{(\lam{g+1}{i_1})^2} +\frac{1}{(\lam{g+1}{i_2})^2}  + \frac{1}{(\lam{g+1}{i_3})^2} = \nonumber \\
&\textstyle~~~~~~~~\left(\frac{1}{\lam{g+1}{i_1}} + \frac{1}{\lam{g+1}{i_2}} + \frac{1}{\lam{g+1}{i_3}}\right)^2 \nonumber \\
&\textstyle~~~~~~~~~~~~~~~~- 2 \left(  \frac{\lam{g+1}{i_1}  + \lam{g+1}{i_2} + \lam{g+1}{i_3} }{\lam{g+1}{i_1}  \cdot \lam{g+1}{i_2} \cdot \lam{g+1}{i_3} } \right). \label{sqrrecexpand.eq}
\end{align}
Then, leveraging the analysis in Zhang et al.~\cite{ZWZZGW10} for $\Sg$, we obtain a recursive expression for the  first term on the right-hand size of (\ref{sqrrecexpand.eq}),
\begin{equation}
\textstyle \left(\frac{1}{\lam{g+1}{i,1}} + \frac{1}{\lam{g+1}{i,2}} + \frac{1}{\lam{g+1}{i,3}}\right)^2 = \frac{(3(v+1))^2}{(\lam{g}{i})^2}. \label{vietresult1.eq}
\end{equation}
For the second term, we use the following equivalences, obtained by application of Vieta's formulae to (\ref{vicrec.eq}),
\begin{align}
&\lam{g+1}{i_1}  \cdot \lam{g+1}{i_2} \cdot \lam{g+1}{i_3} = \lam{i}{g} \\
&\lam{g+1}{i_1}  + \lam{g+1}{i_2} + \lam{g+1}{i_3} = (v+4).
\end{align}
With these equivalences, we can rewrite the second term on the right-hand size of (\ref{sqrrecexpand.eq}) as,
\begin{equation}
 -2\left(\frac{\lam{g+1}{i_1}  + \lam{g+1}{i_2} + \lam{g+1}{i_3} }{\lam{g+1}{i_1} \cdot \lam{g+1}{i_2} \cdot \lam{g+1}{i_3} }\right)  = \frac{-2(\vp+4)}{\lam{i}{g}}. \label{vietresult2.eq}
 \end{equation}

Combining (\ref{vietresult1.eq}) and (\ref{vietresult2.eq}), we obtain to following relationship between $\lam{g}{i}$ and the sum  of the squared inverses of its children,
\begin{equation}
\textstyle \frac{1}{(\lam{g+1}{i_1})^2} +\frac{1}{(\lam{g+1}{i_2})^2}  + \frac{1}{(\lam{g+1}{i_3})^2} = \frac{\left(3(\vp+1)\right)^2}{(\lam{g}{i})^2} - \frac{2(\vp+4)}{\lam{g}{i}}. \label{eiggen.eq}
\end{equation}

We now use the expression (\ref{eiggen.eq}) to find a recursion over the sum of the squared inverses of the eigenvalues of $\Lg$.  As in \cite{ZWZZGW10},
we consider the non-degenerate and degenerate eigenvalues separately.

Recall that $v+1$ is the single, non-zero, non-degenerate eigenvalue of $L_1$.  
Let $\thet{i}$ be the sum of the squared inverses of the $i^{th}$ generation descendants of 
$v+1$, i.e.,
\[
\thet{i} = \sum_{\lambda \in \omegnd{i} - \omegnd{i-1}} \textstyle \frac{1}{\lambda^2}.
\]
Using (\ref{eiggen.eq}), we obtain a recursion over this sum,
\begin{align*}
\thet{i} &= \sum_{\lambda \in \omegnd{i} - \omegnd{i-1}} \textstyle \left(\frac{\left(3(\vp+1)\right)^2}{\lambda^2} - \frac{2(\vp+4)}{\lambda} \right) \\
&= \left(3(\vp+1)\right)^2 \thet{i-1} - 2(\vp+4) \gam{i-1}.
\end{align*}
We then substitute in the expression for $\gam{i-1}$ in (\ref{gamsum1.eq}) and obtain a closed-form expression for $\thet{i}$,
\begin{equation}
\thet{i}  = \textstyle  \frac{7\vp - 12}{9(\vp+2)(\vp+1)^2} 3^{2i} (\vp+1)^{2i} + \frac{2(v+4)}{3\vp+2} 3^{i-1} (\vp+1)^{i-2}. \label{thetafull.eq}
\end{equation}
The full derivation of this expression is given in Appendix~\ref{vicsekrec.app}.

By summing over the the generations $i=0 \ldots g-1$, we can find the sum of the squared inverses of the non-degenerate eigenvalues of $\Lg$.  As we are interested in
the asymptotic behavior for second-order coherence, we only consider the highest order term, which is as follows,
\begin{equation}
\sum_{\lambda \in \omegnd{g}} \frac{1}{\lambda^2} = \sum_{i=0}^{g-1} \thet{i} \sim 3^{2g} (\vp+1)^{2g}. \label{nondegensum2.eq}
\end{equation}

We now consider the descendants of a degenerate eigenvalue of $\Lg$ with value 1. 
Using the recursion (\ref{eiggen.eq}), we can obtain a similar expression for the $i^{th}$ generation descendants of this eigenvalue,
\begin{align*}
\thetdeg{i} &= \left(1 - \textstyle \frac{2(\vp+4)}{3^4(\vp+1)^3(\vp+2)}\right) 3^{2i}(\vp+1)^{2i} \\
&~~~~~~- \textstyle  \frac{2(\vp+4)}{3\vp+2}( 3(\vp+1))^{i-3}.
\end{align*}
The full derivation of this expression is given in Appendix~\ref{vicsekrec.app}.
Then, incorporating the multiplicity of the degenerate eigenvalues as stated in Property \ref{vicsek.prop}, we
can find the sum of the squared inverses of the degenerate eigenvalues, 
\begin{equation}
\sum_{\lambda \in \omegdeg{g}} \frac{1}{\lambda^2}  = \sum_{i=0}^{g-1} \Delta_{g-i} \thetdeg{i} \sim 3^{2g} (\vp+1)^{2g}. \label{degensum2.eq}
\end{equation}

% We can write  similar expression to (\ for the sum of one over the square of the reciprocal of its $i^{th}$ generation descendants.
%Here $\thetdeg{1} = 1$.

Finally, we combine the results in (\ref{nondegensum2.eq}) and (\ref{degensum2.eq}) to find the sum of the squared inverses of the non-zero eigenvalues of $\Lg$, 
\[
\sum_{i=2}^{\Ng} \frac{1}{\left(\lam{g}{i}\right)^2} \sim 3^{2g} (\vp+1)^{2g} = \Ng^{2 + 2 \log(3) / \log(\vp+1)}.
\]

From this result and the definition of second-order network coherence in (\ref{socohdef.eq}), we arrive at the following theorem.
\begin{theorem}
For a generalized Vicsek fractal graph with $N$ nodes, parameterized by the integer $\vp$, the second-order coherence of the system with the dynamics defined in (\ref{do.eq}) is 
\[
\Hs \sim  \frac{1}{\beta^2} N^{1 + 2 \log(3) / \log(\vp+1)} = \frac{1}{\beta^2} N^{1 + (2 / d_f)},
\]
where $d_f$ is the fractal dimension of the graph.
\end{theorem}

\begin{table*}
\footnotesize
\setlength{\extrarowheight}{5pt}
\caption{Examples of self-similar graphs with $N$ nodes and their dimensions and coherence scalings.}
\centering
\begin{tabular}{l|c|c|c}
\textbf{Network} & \textbf{Fractal Dimension} &  $\boldsymbol{\Hf}$ & $\boldsymbol{\Hs}$ \\
\hline 
1-dimensional torus & 1 & $\frac{1}{\beta}N$ & $\frac{1}{\beta^2}N^3$ \\
\hline
Generalized Vicsek Fractal with $\vp = 4$ & $\frac{\log(5)}{\log(3)} \approx 1.46 $ &  $\frac{1}{\beta}N^{\log(3) / \log(5)}$ &  $\frac{1}{\beta^2}N^{1 + 2\left(\log(3) / \log(5)\right)}$  \\
\hline
T Fractal (tree-like fractal with $\tp=1$)&  $\frac{\log(3)}{\log(2)} \approx 1.58$ & $ \frac{1}{\beta}N^{\log(2) / \log(3)}$ & $\frac{1}{\beta^2}N^{1 + 2\left(\log(2) / \log(3)\right)}$ \\
\hline 
Peano Basin Fractal (tree-like fractal with $\tp=2$) & $2$ & $\frac{1}{\beta}\sqrt{N}$   & $\frac{1}{\beta^2}N^2$  \\ 
\hline 
2-dimensional torus & 2 & $\frac{1}{\beta}\log{N}$ & $\frac{1}{\beta^2}N$ \\
\end{tabular}
\label{fractals.tab}
\end{table*}

\section{Coherence and Graph Dimension}
\label{dimension.sec}
The coherence scalings for several families of self-similar graphs are presented in Table \ref{fractals.tab}. 
We note that for the fractal graphs studied in this paper, the coherence expressions for both first-order and second-order systems
exhibit the same scalings a one-dimensional torus or lattice, i.e. coherence scales as $N^{1/d_f}$ for first-order systems and $N^{1+ 2/d_f}$ for second order systems.
This is true even for the fractals with fractal dimension greater than or equal to two.  
These results demonstrate that the fractal dimension does not uniquely determine the scaling behavior of network coherence. The Peano Basin Fractal and the two-dimensional lattice
both have fractal dimension equal to two but exhibit strikingly different coherence scalings.
Therefore, it seems that there other characteristics of the graph that affect the asymptotic scalings of network coherence.

One possible characterization is the spectral dimension, which is defined as follows~\cite{RT83}.
\begin{definition}
Let $\rho(x)$ be the eigenvalue counting function of $L$, i.e., $\rho(x)$ is the number of eigenvalues of $L$ that
have magnitude less than or equal to $x$.  The \emph{spectral dimension} of the graph is 
\[
d_s := 2 \lim_{x \rightarrow \infty} \frac{\log(\rho(x))}{\log(x)}.
\]
\end{definition}

For torus and lattice graphs, the fractal dimension is equal to spectral dimension, and therefore all of the previous coherence results are the same for the fractal and spectral dimensions.
For the families of fractal graphs studied in this paper, the spectral dimension and the fractal dimension are related as $d_s = (2 d_f) / (d_f + 1)$ (see \cite{KL93}).
With this relationship, we can derive expressions for network coherence in terms of the spectral dimension.  
For first order-systems the network coherence is $\Hf \sim N^{2/d_s -1}$, and for second order systems, it is $\Hs \sim N^{4/d_s - 1}$.
We note that the fractals considered in this paper all have spectral dimension less than two.  Therefore, by considering the spectral dimension, we eliminate the conflicting coherence results 
we obtained for graphs with fractal dimension equal to two.  

In general, there is not a straightforward relationship between the fractal and spectral dimensions, and
while the fractal dimension is defined in terms of the graph topology, the relationship between the topology and the spectral dimension
is not generally understood. 
In addition, it is not yet known whether the spectral dimension plays the same role in determining network coherence for families of graphs beyond those
studied in this work. The question of how to generalize our results to other graph structures is an open problem and a subject for future work.

\section{Conclusion}
\label{conclusion.sec}
We have investigated the relationship between the topological dimension of a graph and the 
the per node variance of the deviation from consensus in systems with noisy consensus dynamics.
We have shown that, in first-order systems, the coherence measure is closely related to concepts in electrical networks, random walks, and molecular connectivity.
Drawing directly from literature on random walks in fractal graphs, we  have derived asymptotic expressions for first-order coherence in terms of the network size and 
fractal dimension.  We have then extended this line of analysis to derive asymptotic expressions for second-order coherence for several families of fractal graphs. 
Our analysis shows that the fractal dimension does not uniquely determine the asymptotic behavior of network coherence, and in fact, two self-similar graphs with the same fractal dimension exhibit different coherence scalings.
 We conclude that the question of whether performance of stochastic consensus algorithms in large networks can be captured by purely topological measures, such as the networkÕs fractal dimension, remains open.

\appendices
\section{Derivations for Second-Order Network Coherence in Tree-Like Fractals} \label{tree.app}

First we find the characteristic polynomials for the matrices for the generation 1 graph, $\Pp_1$, $Q_1$ and $R_1$.
We then find the coefficients that we need to solve the recursions (\ref{rectree1r.eq}) - (\ref{rectree2q.eq}). 
For $\Pp_1$, we have,
\[
\Pp_{1}(x) = (-(m+3) + x)(1 - x)^{m+1}.
\]
Note that this is the modified characteristic polynomial defined in (\ref{charmod.eq}).
The coefficient corresponding to the second order terms (in $x$) is
\begin{equation} \label{p12.eq}
p_1^{(2)} = -(m+1)  - \textstyle \frac{(m+3)(m+1)m}{2}.
\end{equation}
For $Q_1$, the characteristic polynomial is
\[
Q_1(x) = (1 + (-m-3)x + x^2)(1 - x)^m. 
\]
The coefficient for the second order term (in $x$) is
\begin{equation} \label{q12.eq}
q_1^{(2)} = \frac{3}{2}m^2 + \frac{5}{2}m + 1.
\end{equation}
For $R_1$, the characteristic polynomial is
\[
R_1(x) = (2 + (-m-3)x + x^2)(1-x)^{m-1}.
\]
The coefficients for the first order term (in $x$) is
\begin{equation} \label{r11.eq}
r_1^{(1)} = -3m - 1.
\end{equation}

We now solve the recursion equation for $\rgo$.
Substituting the expressions for $\pgz$, $\qgz$, $\rgz$, and $\qgo$ from \cite{LBZ10} into (\ref{rectree1r.eq}), we obtain,
\begin{align*}
r_{g+1}^{(1)} =&~2\rgo \\
&~-2^{g+1}(m+1)(m+2)^{g-1}[ 1 - 2^g + 2^g(m+2)] \\
&~~+ (m+1)2^{2g} + m 2^{2g}(-(m+2)^g - 1) \\
 =~& 2 \rgo +  \textstyle \frac{-2m^2 -5m -2}{(m+2)} 2^{2g} (m+2)^g \\
 &~~- \textstyle \frac{2(m+1)}{(m+2)} 2^g (m+2)^g + 2^{2g}.
\end{align*}
Then, using the value for $r_1^{(1)}$ in (\ref{r11.eq}), we obtain a closed-form expression for the above recursion,
\begin{eqnarray*}
\rgo &=& 2^{g-1} r_1^{(1)} + \textstyle \frac{(-2m^2 -5m -2)}{(m+2)}2^g \displaystyle \sum_{i=1}^{g-1}(2(m+2))^i \\
&& -  \textstyle \frac{2(m+1)}{(m+2)} 2^g \displaystyle \sum_{i=1}^{g-1} (m+2)^i ~+~ 2^g \sum_{i=1}^{g-1} 2^i \\
&=& 2(-3m-1) \textstyle - \frac{2m^2 + 5m + 2}{2m+3} 2^{2g}(m+2)^{g-1} \\
&& \textstyle + \frac{4m^2 + 10m + 4}{2m+3} 2^g~-~ 2^{g+1}(m+2)^{g-1} + 2^{2g}.
 \end{eqnarray*}
The highest order term of $\rgo$ is thus,
\[
\rgo \sim -2^{2g} (m+2)^g.
\]

Next, we solve the recursion for $\qgt$, which is given by,
\begin{equation} \label{qrec.eq}
q_{g+1}^{(2)} = (m+2) \qgt + D_g,
\end{equation}
where 
\begin{align*}
D_g &= \textstyle \frac{(m+1)(m+2)}{2}[\qgo]^2 \\
&~~~~- ((m^2 + 2m + 2)(m+2)^g +1)2^g \qgo \\
&~~~~+ 2^g (m+1) \pgo  - (m+1)(m+2)^g \rgo. 
\end{align*}
Note that the closed-form expressions for $\pgo$ and $\qgo$ are given in \cite{LBZ10}.
%Some useful simplifications
%\[
%\qgo \sim -2^g(m+2)^g
%\]
%\[
%\pgo \sim 2^g (m+2)^{2g}
%\]
%\[
%(\qgo)^2 \sim 2^{2g} (m+2)^{2g}
%\]
%So the highest order term of $D_g$ is of order
%\[
%D_g \sim 2^{2g}(m+2)^{2g}
%\]
Solving the recursion (\ref{qrec.eq}), using the initial value of $q_1^{(2)}$ in (\ref{q12.eq}), we find the following closed-form expression,
\begin{eqnarray*}
\qgt &\sim&  \textstyle (m+2)^{g-1}\left(\frac{3}{2}m^2 + \frac{5}{2}m + 1\right)  \\
&&~~+ (m+2)^g \sum_{i=1}^{g-1} (4(m+2))^i.
\end{eqnarray*}
The highest order term of $\qgt$ is thus,
\[
\qgt \sim  2^{2g} (m+2)^{2g}
\]

Finally, we consider the recursion for $\pgt$,  
\begin{equation} \label{prec.eq}
p_{g+1}^{(2)} = (m+2)\pgt+ E_g,
\end{equation}
where 
\begin{eqnarray*}
E_g&=& (m+1)(m+2)(-(m+2)^g - 1)\qgt \\
&&+\textstyle \frac{m(m+1)(m+2)}{2}(-(m+2)^g-1)[\qgo]^2  \\
&&+ (m+1)(m+2)\pgo \qgo  \\
&&+(m+1)(m+2)\qgt +\textstyle \frac{(m+1)^2(m+2)}{2}[\qgo]^2. 
\end{eqnarray*}
Using the initial value of $p_1^{(2)}$ in (\ref{p12.eq}), we obtain a closed-form expression for $\pgt$,
\[
\pgt = (m+2)^{g-1} p_1^{(2)} +  \textstyle \sum_{i=1}^{g-1}(m+2)^{g-i}E_i.
\]
The highest order for of $\pgt$ is thus,
\[
\pgt \sim - 2^{2g}(m+2)^{3g}.
\]

We can now find the asymptotic order of the desired sum in (\ref{doco.eq}),
\[
\sum_{i=2}^{\Ng} \frac{1}{(\lambda_i)^2} = \textstyle \left( \frac{\pgo}{\pgz}\right)^2 - 2\frac{\pgt}{\pgz} \sim 2^{2g}(m+2)^{2g}.
\]

\section{Derivations for Second-Order Network Coherence in Vicsek Fractals} \label{vicsekrec.app}

Here we show the details of the derivation of $\thet{i}$ in (\ref{thetafull.eq}).
The recursive expression for $\thet{i}$ is,
\begin{align*}
\thet{i} &= \sum_{\lambda \in \omegnd{i} - \omegnd{i-1}} \textstyle \left(\frac{\left(3(\vp+1)\right)^2}{\lambda^2} - \frac{2(\vp+4)}{\lambda} \right) \\
&= \left(3(\vp+1)\right)^2 \thet{i-1} - 2(\vp+4) \gam{i-1}.
\end{align*}
We solve the above recursion to obtain,
\begin{align} 
 \thet{i}  &= \left(3(\vp+1)\right)^{2i} \thet{0} \nonumber \\
 &~~~ - 2(\vp+4) \sum_{j=0}^{i-1} \left(3(\vp+1)\right)^{2(i-j-1)} \gam{j}.  \label{thetaclosed.eq}
\end{align}
 
Recall that $\vp+1$ is single, non-degenerate, non-zero eigenvalue of $L_1$.  Therefore, $\thet{0} = \frac{1}{(\vp+1)^2}$.
Substituting this and the value for $\gam{j}$ given in (\ref{gamsum1.eq}), we simplify (\ref{thetaclosed.eq}) as follows,
\begin{align*}
\thet{i} & = \left(3(\vp+1)\right)^{2i} \textstyle \frac{1}{(\vp+1)^2} \\
&~~ \textstyle -\frac{2(v+4)}{v+1} \displaystyle \sum_{j=0}^{i-1}  \left(3(\vp+1)\right)^{2(i-j-1)} (3(\vp+1))^{j})\\
& = 3^{2i} (\vp+1)^{2i-2} \\
&~~- 2   (v+4) 3^{2(i-1)} (v+1)^{2i-3} \sum_{j=0}^{i-1} (3(\vp+1))^{-j} \\
& = 3^{2i} (\vp+1)^{2i-2} \\
&~~~~-\textstyle \frac{2(v+4)}{3v+2} \left(3^{2i-1)} (\vp+1)^{2i-2} - 3^{i-1} (\vp+1)^{i-2} \right) \\
& = \textstyle  \frac{7\vp - 12}{9(\vp+2)(\vp+1)^2} 3^{2i} (\vp+1)^{2i} + \frac{2(v+4)}{3\vp+2} 3^{i-1} (\vp+1)^{i-2}.
\end{align*}

For the degenerate eigenvalue of 1 of $L_1$ the closed-form expression for the recursion over its descendants is the same as for the non-degenerate eigenvalue $(v+1)$,
\begin{align} 
 \thetdeg{i}  &= \left(3(\vp+1)\right)^{2i} \thetdeg{0} \nonumber \\
&~~~ - 2(\vp+4) \sum_{j=0}^{i-1} \left(3(\vp+1)\right)^{2(i-j-1)} \gamdeg{j}. \label{degenclosed.eq}
\end{align}
In this case, $\thetdeg{0} = 1$. The closed-form expression for $\gamdeg{j}$ is given in (\ref{degenclosed.eq}).
\begin{align*}
\thetdeg{i} & = \left(3(\vp+1)\right)^{2i} \\
&~~~~ - 2(v+4)  \sum_{j=0}^{i-1}  \left(3(\vp+1)\right)^{2(i-j-1)} (3(\vp+1))^{j}) \\
&=  \left(3(\vp+1)\right)^{2i} \\
&~~~~- \textstyle \frac{2(\vp+4)}{3\vp+2} \left((3(\vp+1))^{2i-3} - (3(\vp+1))^{i-3}\right) \\
&= \left(1 - \textstyle \frac{2(\vp+4)}{3^4(\vp+1)^3(\vp+2)}\right) 3^{2i}(\vp+1)^{2i} \\
&~~~~~~ \textstyle -  \frac{2(\vp+4)}{3\vp+2}( 3(\vp+1))^{i-3}.
\end{align*}

\section*{Acknowledgment}
The authors would like to thank Stephen Boyd for a useful insight and discussion regarding networks with fractional Hausdorff dimension. 

% Can use something like this to put references on a page
% by themselves when using endfloat and the captionsoff option.
%\ifCLASSOPTIONcaptionsoff
%  \newpage
%\fi

% trigger a \newpage just before the given reference
% number - used to balance the columns on the last page
% adjust value as needed - may need to be readjusted if
% the document is modified later
%\IEEEtriggeratref{8}
% The "triggered" command can be changed if desired:
%\IEEEtriggercmd{\enlargethispage{-5in}}

% references section

% can use a bibliography generated by BibTeX as a .bbl file
% BibTeX documentation can be easily obtained at:
% http://www.ctan.org/tex-archive/biblio/bibtex/contrib/doc/
% The IEEEtran BibTeX style support page is at:
% http://www.michaelshell.org/tex/ieeetran/bibtex/
%\bibliographystyle{IEEEtran}
% argument is your BibTeX string definitions and bibliography database(s)
%\bibliography{IEEEabrv,../bib/paper}
%
% <OR> manually copy in the resultant .bbl file
% set second argument of \begin to the number of references
% (used to reserve space for the reference number labels box)
 \bibliographystyle{IEEEtran} 
 \bibliography{IEEEabrv,fractals}

\end{document}